\documentclass[paper]{JHEP}
\usepackage{amsmath}
\usepackage{epsfig}

\def\al{\alpha}
\def\as{\alpha_{\mbox{\scriptsize s}}}
\def\aef{\alpha_{\mbox{\scriptsize eff}}}

\def\qq{q\bar{q}}
\def\ee{e^+e^-}

\def\MSbar{\overline{\mbox{MS}}}
\def\MSbar{\overline{\mbox{\scriptsize MS}}}
\def\GeV{\mathop{\rm Ge\!V}}

\def\be{\beta}

\def\eps{\epsilon}
\def\de{\delta}

\def\om{\omega}
\def\gam{\gamma}
\def\lam{\lambda}

\def\mat{\mbox{\scriptsize mat}}

\def\exact{\mbox{\scriptsize exact}}

\def\cO#1{{\cal{O}}\left(#1\right)}
\def\half{\mbox{\small $\frac{1}{2}$}}

\def\abs#1{\left| \: #1 \: \right|}

\def\VEV#1{\left\langle#1\right\rangle}

\def\PT{\mbox{\scriptsize PT}}
\def\NP{\mbox{\scriptsize NP}}

\def\bnu{\bar{\nu}}

\def\ka{\kappa}

\def\cD{{\cal{D}}}

\def\cM{{\cal{M}}}
\def\cR{{\cal{R}}}

 \newskip\humongous \humongous=0pt plus 1000pt minus 1000pt
   \newif\ifdtup

\def\la{\mathrel{\mathpalette\fun <}}

\def\fun#1#2{\lower3.6pt\vbox{\baselineskip0pt\lineskip.9pt
  \ialign{$\mathsurround=0pt#1\hfil##\hfil$\crcr#2\crcr\sim\crcr}}}

\title{QCD analysis of $D$-parameter\\ in near-to-planar three-jet
events}

\author{ A.~Banfi,\\ Dipartimento di Fisica, Universit{\`a} di
Milano--Bicocca and INFN, Sezione di Milano, Italy} 

\author{
Yu.L.~Dokshitzer,\\ LPT, Universit\'e de Paris XI, Centre d'Orsay,
France \footnote{on leave from PNPI, Gatchina, St.~Petersburg,
188350, Russia}}

\author{ G.~Marchesini,\\ Dipartimento di Fisica, Universit{\`a} di
Milano--Bicocca and INFN, Sezione di Milano, Italy}

\author{ G.~Zanderighi.\\ Dipartimento di Fisica Nucleare e Teorica,
  Universit{\`a} di Pavia and INFN, Sezione di Pavia, Italy}

\abstract{ We present the QCD analysis of $D$-parameter distribution
  in near-to-planar 3-jet $\ee$ annihilation events. We derive the
  all-order resummed perturbative prediction and the leading power
  suppressed non-perturbative corrections both to the mean value and
  the distribution.  Here non-perturbative corrections are larger than
  in $2$-jet shape observables, so that higher order non-perturbative
  effects could be relevant. Experimental data (not yet available) are
  needed in order to cast light on this important point.  The
  technique we develop aims at improving the accuracy of the
  theoretical description of multi-jet ensembles, in particular in
  hadron-hadron collisions.  }

\keywords{QCD, Jets, LEP and SLC Physics, Nonperturbative Effects}

\preprint{
     Bicocca--FT--01/10\\
     LPT--Orsay--01/37\\
     Pavia--FNT/T--01/10\\
     hep-ph/0104162\\
     April 2001}

\begin{document}

\section{Introduction}
The analysis of event shape distributions in $\ee$ has provided
various tests of QCD \cite{Exp-shape} and measurements of the running
coupling \cite{Exp-running}.
The shape observables which have been most intensively studied and
tested are: thrust $T$, $C$-parameter, jet mass $M^2$ and broadening
$B$.  Their distributions are collinear and infrared safe (CIS) and
therefore can be computed order-by-order in perturbation (PT) theory.
Specially interesting is the region of two narrow jets
($1\!-\!T,C,M^2/Q^2,B\ll 1$), where the PT expansion needs to be
resummed so that the QCD structure is most intensively probed. The
available PT results \cite{PTstandard} for these typical $2$-jet
observables\footnote{Sometime these observables are called $3$-jet
observables since the first contribution involves three particles in
the final state. We prefer to call them $2$-jet observables since we
are interested to the kinematical region in which one particle of the
three is soft.} involve all-order resummation of double- (DL) and
single-logarithmic (SL) contributions and matching of the approximate
resummed expressions with the exact second order matrix elements.

To make quantitative predictions one needs to go beyond PT
calculations and to take into account the $1/Q$--suppressed power
corrections arising from the interaction in the confinement region.
It has been proposed \cite{NPgeneral} that these non-perturbative (NP)
corrections can be estimated by extrapolating the running coupling
into the large distance region.  A systematic method for this
extrapolation is provided by the dispersive approach \cite{DMW}.  The
leading NP corrections to the shape distributions involve a single
parameter, usually denoted by $\al_0$, which is given by the integral
of the QCD coupling over the region of small momenta $k\le \mu_I$
(with the infrared scale $\mu_I$ conventionally chosen to be
$\mu_I=2\GeV$).
They have been computed at two-loop level to take into account effects
of non-inclusiveness of jet observables \cite{Milan}.  Such a
procedure for computing the $1/Q$--power corrections is consistent
with the data and the NP parameter $\al_0$ has been measured and
appears to be universal with a reasonable accuracy~\cite{Exp-shape}.

We have recently studied the distribution in the thrust minor $T_m$, a
CIS shape observable characterising $3$-jet events.  It starts at
order $\as^2$ and measures the radiation out of the event plane in
$e^+e^-$ annihilation. The event plane is defined by the $T$ and $T_M$
axes, with $T_M$ is the thrust major.  The most interesting region in
which one probes QCD is that of nearly planar $3$-jet events, $T_m\ll
T_M\sim T$, where PT resummation is needed.  The methods developed for
the analysis of $2$-jet observables $T,C,M^2/Q^2$ and $B$ were
extended to the case of~$T_m$ in~\cite{Kout}.

In this paper we report the analysis of the distribution in another
typical $3$-jet observable, the $D$-parameter which has been
introduced in \cite{ERT}. This is a CIS shape observable which, as
$T_m$, measures the radiation out of the event plane.  In the nearly
planar $3$-jet region ($D\ll1$), the main difference between $D$ and
$T_m$ is that only soft particles at large angles with respect to the
event plane, $p_{\mbox{\scriptsize out}} \sim p_{\mbox{\scriptsize
in}}\propto D\ll 1$, contribute to $D$, contrary to the $T_m$
case. The fact that energetic hadrons with $p_{\mbox{\scriptsize
in}}\gg p_{\mbox{\scriptsize out}}$ do contribute to $T_m$, gives
rise, in particular, to additional logarithmic enhancement of the NP
contribution from the regions where a small transverse momentum
energetic gluon is emitted close to the direction of one of the three
jets.  This difference is similar to that between $B$ and
$T,C,M^2/Q^2$ in $2$-jet events \cite{Milan}: an energetic hadron
emitted near the thrust axis contributes to $B$ but does not
contribute to the other shape observables.

Our study is performed with the accuracy necessary to make
quantitative predictions. We perform the all-order resummation of DL-
and SL-enhanced PT contributions, match it \cite{PTstandard} with the
exact fixed order matrix element calculation and compute the leading
$1/Q$ NP correction at the two-loop order.  Actually, the fact that
only hadrons at large angles contribute to $D\ll1$ makes the present
analysis significantly simpler than that of $T_m$.

We find that, as in the $T_m$ case, the structure of the result is
quite rich, especially for small $D$ where both the PT and NP
components of the $D$ distribution essentially depend on the geometry
of the event (the angles between jets).
%%%
Not only will the analysis of the $D$ distribution provide an
alternative measurement of the QCD coupling $\as$.  It gives a
powerful tool for accessing genuine confinement hadronisation effects,
for extracting the NP parameter $\al_0$ and testing its universality.

The paper is organised as follows.
In section~2 we introduce the observable and discuss the
characteristic kinematics in the near-to-planar $3$-jet region.
In section~3 we discuss the resummation procedure.
In section~4 we analyse the PT resummation at SL level.
In section~5 we analyse the leading NP corrections in terms of the
universal parameter $\al_0$.
In section~6 we report the final result and discuss the necessary
ingredients of the one-loop matching.
In section~7 we report the numerical analysis of the distribution and
the mean.
In section~8 we summarise and discuss the results.

\section{The observable \label{sec:Observable}}
The $D$ parameter is defined as \cite{ERT}
\begin{equation}
  \label{eq:Dpar}
  D \equiv 27 \>\mbox{det}\> \theta\!=\!27\lam_1\lam_2\lam_3\>,\qquad
  \theta_{\al\be}=
\frac{1}{\sum_h\abs{\vec p_h}}
\sum_h \frac{p_{h\al}\,p_{h\be}}{\abs{\vec p_h}}\>,
\end{equation}
with $\vec{p}_h$ the momentum of the emitted hadron $h$ and $p_{h\al}$
its $\al$-component ($\al=1,2,3$). The eigenvalues $\lam_i$ satisfy
the conditions $0\le\lam_i\le1$ and $\sum_i\lam_i=1$. We assume the
order $\lam_3\le\lam_2\le\lam_1$.

To select near-to-planar events we introduce a lower limit $y_c$ for
the $3$-jet resolution variable $y_3$, defined\footnote{Given the set
of all momenta, one defines the ``jettiness'' variable \cite{jettin}
of any pair $p_{h}$ and $p_{h'}$ as the quantity
$y_{hh'}=2\,(1-\cos\theta_{hh'})\,\min(E_h^2, E_{h'}^2)/Q^2$.  The
pair of momenta $p_{\bar h},p_{\bar h'}$ with the minimum distance
$y_3=y_{\bar h, \bar h'}=\min_{hh'} y_{hh'}$ are substituted with the
pseudoparticle (jet) momentum $p_{h''}=p_{\bar h}+p_{\bar h'}$.  The
procedure is repeated with the new momentum set till only three jets
are left. Then the final value of $y_3$ is defined as the three-jet
resolution of the event.}  according to the $k_{T}$ (Durham) algorithm
\cite{Durham}.  We study the following normalised integrated
distribution
\begin{equation}
\label{eq:Sigma}
\begin{split}
&\Sigma(D,y_c)=\frac{1}{\sigma(y_c)}\sum_m\!\int\! {d\sigma_m}\>
\Theta(y_3-y_c)\>\Theta\left(D\!-\!27\,\mbox{det}\,\theta\right)\>,\\
&\sigma(y_c)=\sum_m\!\int\! {d\sigma_m}\>\Theta(y_3-y_c)\>,
\end{split}
\end{equation}
where $d\sigma_m$ denotes the differential distribution in the $m$
final particles with momenta~$p_h$.

We comment here on the fact that we use the variable $y_3$ to select
the $3$-jet events (instead of, for instance, the thrust $T$), which
makes it easier to interpret the $1/Q$ power corrections that, as we
shall see, are present in the D-distribution.
%%%
Indeed, the difference between the hadron and parton level values of
the thrust variable $T$ is known to be of the order of $1/Q$, while
the values of $y_3$ for hadrons and partons is not affected by $1/Q$
corrections \cite{DMW}.  Therefore, the $1/Q$ contribution to the
distribution \eqref{eq:Sigma} can be looked upon as a genuine NP
correction to the variable $D$.  On the contrary, substituting $T<T_c$
for $y_3>y_c$ in the distribution \eqref{eq:Sigma}, the $1/Q$
correction would have to be shared between $D$ and $T$.

The region of small $D$ and relatively large $y_c$ (we consider
typical values of $y_c$ in the range $0.025-0.1$) corresponds to the
region
\begin{equation}
  \label{eq:3jet}
\lam_3\ll\lam_2\la\lam_1\>, \qquad T_m\ll T_M \la T\>.
\end{equation}

Here $\lam_1,\lam_2,y_3$ and $T,T_M$ are determined by the hard
momenta characterising the three jets. The smallest eigenvalue
$\lam_3$ and $T_m$ are determined by the out-of-event-plane soft
momentum component of the particles around and between the jets
(inter- and intra-jet radiation).
Taking the plane of the event as the $\{y,z\}$-plane, the dominant
contribution to $\lam_3$ in the region \eqref{eq:3jet} is given~by
\begin{equation}
  \label{eq:lam3def}
 \lam_3\>\simeq \sum_{h}\frac{p^2_{h\,x}}{E_hQ}\,.
\end{equation}
Due to the energy factor $E_h$ in the denominator, $\lam_3$ gets the
leading contribution from hadrons with $E_h\sim p_{h\,x}$.  Instead,
the observable $T_m$
\begin{equation}
  \label{eq:Kout}
  T_m=\sum_{h}\frac{|p_{hx}|}{Q}\>,
\end{equation}
as mentioned in the Introduction, receives contributions from
particles with arbitrary large energies, and in particular those that
are quasi-collinear with one of the hard jets.

\section{Parton process and resummation \label{sec:Parton}}
Our first aim is to express the distribution $\Sigma(D,y_c)$ in terms
of parton processes in the near-to-planar region \eqref{eq:3jet}.
Here the parton events can be treated as $3$-jet events generated by a
hard quark-antiquark-gluon system accompanied by an ensemble of
secondary partons $k_i$. In this region, the quantities
$\lam_1,\,\lam_2,\,T,\,T_M$ and the $3$-jet resolution variable $y_3$
are determined by the three momenta of the hard quark, antiquark and
gluon which we denote by $P_1,P_2$ and $P_3$, respectively.
Introducing the Born $x$-invariants (for the Born Kinematics see
Appendix ~\ref{App:Born})
\begin{equation}
  \label{eq:xa}
x_a\equiv \frac{2P_aQ}{Q^2}\>, \qquad 2=x_1+x_2+x_3\>,
\end{equation}
we have
\begin{equation}
\label{eq:lam12}
\begin{split}   
&\lam_1\, \lam_2=\frac{2(1-x_1)(1-x_2)(1-x_3)}{x_1x_2x_3}\>,
\qquad  y_3=\frac{x_{\min}(1-x_{\max})}{2-x_{\max}-x_{\min}}\>,
\end{split}
\end{equation}
with 
\begin{equation}
  \label{eq:xmax}
x_{\max}=\max \{x_1,x_2,x_3\}\>,\quad x_{\min}=\min\{x_1,x_2,x_3\}\>.
\end{equation}
At the Born level, all three partons are in the event plane so that
$D=\lambda_3=0$.  With account of secondary partons $k_i$, radiation
out of the event plane is generated and one gets $\lam_3>0$.  At the
same time, the hard quark, antiquark and gluon momenta acquire recoils
and move out of the event plane.  In the near-to-planar region
\eqref{eq:3jet}, both the hard parton recoil and the secondary parton
momenta can be treated as small (soft)\footnote{As we have shown in
\cite{Kout}, finite rescaling of the in-plane momenta, due to hard
collinear splittings, gets absorbed into the first hard correction to
the emission probability of soft gluons, which is then resummed and
embodied into the radiator.  Bearing this in mind, all hard parton
recoils and the secondary parton momenta $k_i$ can be effectively
treated as small.  }. To leading order the smallest eigenvalue
$\lam_3$ is given by
\begin{equation}
  \label{eq:lam3} 
\lam_3\>=\> \sum_{i}\frac{k^2_{i\,x}}{\om_iQ}\,,
\end{equation}
with $k_{i\,x}$ the out-of-event plane component and $\om_i$ the
energy of a secondary soft parton $i$.  Corrections are quadratic in
the ``soft'' parameter $k_x/Q\sim D\ll1$.  In particular, due to the
presence of the energy in the denominator, the contributions from the
recoiling hard primary partons $\qq,g$ are of second order
($\cO{D^2}$) and have been neglected in \eqref{eq:lam3}.  The
quantities $\lam_1\lam_2$ and $y_3$ are given, to leading order, by
\eqref{eq:lam12} with corrections linear and quadratic in $D$
respectively.

\subsection{Resummation of soft radiation}
The starting point for the parton resummation of the distribution
\eqref{eq:Sigma} is the factorisation of soft emission from the hard
parton system.  The distribution $M^2_{n}$ for the emission of $n$
soft partons from the primary $\qq,g$ system can be factorised in the
form
\begin{equation}
  \label{eq:soft}
  M^2_{n}\,(P_a,k_1\ldots k_n)\simeq
  M^2_{0}\,(P_a)\cdot S_{n}\,(P_a,k_1\ldots k_n)\>,
\end{equation}
where we have used the fact that the hard parton recoils can be
neglected and the actual $\qq g$ momenta replaced by their Born values
$P_a$.  The first factor is the squared first order matrix element
giving the Born distribution
\begin{equation}
  \label{eq:sigmaBorn}
  \frac{d\sigma^{(0)}}{dx_1dx_2}=
\frac{C_F\as}{2\pi}\>\frac{x_1^2+x_2^2}{(1-x_1)(1-x_2)}\>,
\quad \as=\as(Q)\>,
\end{equation}
where, from now on, $x_1$ and $x_2$ mark the quark and antiquark
invariant energy fractions (or vice versa), and $x_3=2-x_1-x_2$ is
that of the hard gluon.

The second factor $S_{n}$ describes the distribution for emitting $n$
soft partons $k_i$ from the hard $\qq,g$ system.  This distribution
can also be factorised into the product of independent soft emissions.
The structure of this factorisation depends on the required accuracy.
Since we aim at SL accuracy, we follow the analysis of \cite{Kout} in
which $S_{n}$ is factorised at two-loop order.

In the region \eqref{eq:3jet}, from the factorised expression 
\eqref{eq:soft} at parton level we can write 
\begin{equation}
\label{eq:Sigma-fact}
\begin{split}
&\Sigma(D,y_c)\simeq \frac{1}{\sigma(y_c)}
\int dx_1dx_2\,\Theta(y_3-y_c)
\left\{
C(\as)\cdot 
\frac{d\sigma^{(0)}}{dx_1dx_2}\cdot S(D,x_1,x_2))
\right\},\\
&\sigma(y_c)\simeq
\int dx_1dx_2\,\Theta(y_3-y_c)\>\frac{d\sigma^{(0)}}{dx_1dx_2}\>,
\end{split}
\end{equation}
with $y_3$ given by the Born expression \eqref{eq:lam12}.  The first
factor in the curly brackets is the non-logarithmic coefficient
$C(\as)=1+\cO{\as}$, which is, in general, a function of $y_3$ and
$D$, and has a finite $D\to 0$ limit.  The soft factor $S(D,x_1,x_2)$
accumulates all logarithmic dependences on $D$, and is given by
\begin{equation}
\label{eq:Sdef}
\begin{split}
S(D,x_1,x_2)\>=\>\sum_n \frac1{n!} \int 
\prod_{i}^n \frac{d^3k_i}{\pi \om_i}\> S_{n}(P_a,k_1\ldots k_n)
\>\Theta\! \left(\!\lam_3\!-\!\sum_i\frac{k^2_{ix}}{\om_iQ}\right),
\end{split}
\end{equation}
(with $\lam_3={D}/{27\lam_1\lam_2}$).  Taking the factorised structure
of $S_{n}$ at two loops, this expression for $S(D,x_1,x_2)$ is
accurate to SL level, see \cite{Kout}.

Given the factorised expression for $S_{n}$, in order to sum up the
series in \eqref{eq:Sdef} in suffices to factorise the theta-function
constraint by using the Mellin representation. We obtain
\begin{equation}
  \label{eq:S}
  S(D,x_1,x_2)=
\int \frac{d\nu}{2\pi i \nu}\>e^{\nu \lam_3}\>\sigma(\nu,x_1,x_2)\>,
\end{equation}
where 
\begin{equation}
  \label{eq:sigma}
  \sigma(\nu,x_1,x_2)=\sum_n \frac1{n!} \int 
\prod_{i}^n \frac{d^3k_i}{\pi \om_i}\> 
e^{-\nu\frac{k^2_{ix}}{\om_iQ}}\cdot S_{n}(P_a,k_1\ldots k_n)
\>\equiv\> e^{-\cR(\nu,x_1,x_2)}.
\end{equation}
The contour in \eqref{eq:S} runs parallel to the imaginary axis with
Re $\nu>0$.  The near-to-planar region $D\ll1$ corresponds to the
region of large Mellin variable $\nu\gg1$.

We show that the radiator is given by a PT contribution and a NP
correction
\begin{equation}
  \label{eq:Rad}
  \cR(\nu,x_1,x_2)=
\cR^{\PT}(\nu,x_1,x_2)+\de\cR(\nu,x_1,x_2)\>.
\end{equation}
In the next section we compute the PT contribution at SL level.  In
section~\ref{sec:NP} we compute $\de\cR$, the leading $1/Q$ correction
including the effect of non-inclusiveness of the $D$-parameter (Milan
factor).

\section{PT contribution  at SL accuracy \label{sec:PT}}
To obtain the PT radiator $\cR^{\PT}(\nu,x_1,x_2)$ we follow the
procedure described in detail in \cite{Kout} for the calculation of
the $T_m$ distribution.  The present case is simpler since the hard
parton recoil can be neglected.  At SL level we have
\begin{equation}
  \label{eq:RadPT1}
\begin{split}  
\cR^{\PT}=\frac{N_c}{2}\left(r_{13}+r_{23}-\frac{1}{N_c^2}\,r_{12}\right),
\end{split}
\end{equation}
where
\begin{equation}
  \label{eq:rab}
r_{ab}(\nu,x_1,x_2)=\int\frac{d^3k}{\pi\om}w_{ab}(k)\,
\left[1-e^{-\nu\frac{k_x^2}{\om Q}}\right],
\qquad w_{ab}(k)=\frac{\as(k^2_{t,ab})}{\pi k^2_{t,ab}}\>,  
\end{equation}
with $w_{ab}(k)$ the soft distribution for the dipole $ab$. Here the
running coupling is defined in the physical scheme \cite{CMW} and
$k_{t,\,ab}$ is the invariant transverse momentum of $k$ with respect
to the hard parton pair $P_a,P_b$, defined as
\begin{equation}
  \label{eq:kabt}
k^2_{t,ab} =\frac{2(P_ak)(kP_b)}{(P_aP_b)}\>.
\end{equation}
The unity in the square bracket in \eqref{eq:rab} takes into account
the virtual corrections. The contribution from the source $u(k)=
e^{-\nu{k_x^2}/{\om Q}}$ results from the exponentiation of real
emissions, see \eqref{eq:sigma}.  To reach SL accuracy one needs to
take into account also the correction coming from hard collinear
parton splittings, which will be embodied into a redefinition of the
hard scales, see later.

\subsection{Explicit calculation of the PT radiator}
In the laboratory frame the hard parton momenta $P_a$ and $P_b$ are
not back-to-back and the expression of soft dipole distribution
$w_{ab}(k)$ is rather cumbersome (see \eqref{eq:kabt}).  It is then
natural to evaluate $r_{ab}$ starting from its expression in the
centre-of-mass frame of the dipole $P_a+P_b$,
\begin{equation}
  \label{eq:cmab}
  P^*_a\!=\!\frac{Q_{ab}}{2}\>(1,0,0,1)\,,\quad
  P^*_b\!=\!\frac{Q_{ab}}{2}\>(1,0,0,-1)\,,\quad 
Q^2_{ab}\!=\!2P_aP_b\!=\!Q^2(x_a\!+\!x_b-1)\,,
\end{equation}
in which the distribution $w_{ab}(k)$ and the phase space are
given by
\begin{equation}
  \label{eq:wabcm}
  w_{ab}(k)=\frac{\as(\ka^2)}{\pi\ka^2}\>,\qquad 
\frac{d^3k}{\pi\om}=d\ka^2\,\frac{d\phi}{2\pi}d\eta\>,
\end{equation}
with $\ka^2$, $\eta$ and $\phi$ the squared transverse momentum, the
rapidity and the azimuthal angle of the soft gluon, respectively.
To compute the dipole radiator $r_{ab}(\nu)$ we need to express the
source for our observable $u(k)=e^{-\nu{k^2_{x}}/{\om Q}}$ in the
frame \eqref{eq:cmab}. While the variable $k_x$ is the same in the two
frames, $ k_{x}=\ka\sin\phi$, the gluon energy $\om$ in the laboratory
frame (see \eqref{eq:Pa} in Appendix~\ref{App:Born}) is given, in
terms of the dipole c.m.s. variables
\begin{equation}
  \label{eq:om}
  \om=\ka\left(
A_{ab}\cosh(\eta\!+\!\eta^0_{ab})-\sqrt{A_{ab}^2\!-\!1}\>\cos\phi\right)\>,
\end{equation}
with $A_{ab}$ and $\eta^0_{ab}$ the following functions of $x_a,x_b$ 
\begin{equation}
\label{eq:A}
\begin{split}  
& A_{ab}=\frac{Q}{Q_{ab}}\sqrt{x_ax_b}
=\sqrt{\frac{x_ax_b}{x_a+x_b-1}}\>,
\qquad 
\tanh\eta^0_{ab}
=\frac{x_a\!-\!x_b}{x_a\!+\!x_b}\>.
\end{split}
\end{equation}
Finally, using the frame \eqref{eq:cmab}, the $r_{ab}$ radiator can be
written in the form
\begin{equation}
  \label{eq:rabcm}
  r_{ab}(\nu,x_1,x_2)=\int_0^{Q^2}\frac{d\ka^2}{\ka^2}\frac{\as(\ka^2)}{\pi}
\int_{-\pi}^{\pi}\frac{d\phi}{2\pi}
\int_{-\eta_M}^{\eta_M}d\eta\left(1\!-\!e^{-\nu\frac{\ka}{Q}\,\tau}\right),
\quad \eta_M=\ln \frac{Q_{ab}}{\ka}\>,
\end{equation}
where $\tau$ is the following function of $\eta$ and $\phi$
\begin{equation}
  \label{eq:tau}
\tau=\frac{\sin^2\phi}
{A_{ab}\cosh(\eta\!+\!\eta^0_{ab})-\sqrt{A_{ab}^2\!-\!1}\,\cos\phi}\>.
\end{equation}
Note that, since $Q_{ab}\sim Q$, to SL accuracy, we do not care for
the precise upper limit in $\ka$ as long as it is of order $Q$.

The calculation of $r_{ab}$ is performed in Appendix \ref{App:PT} and
one finds, to SL accuracy,
\begin{equation}
\label{eq:rab-fine}
r_{ab}(\nu,x_1,x_2)
= 2\int_{{Q^2}/{\nu}}^{Q^2} \frac{d\ka^2}{\ka^2}\frac{\as(\ka^2)}{\pi}
  \ln\frac{Q_{ab}}{\ka}
+2\int_{{Q^2}/{\nu^2}}^{{Q^2}/{\nu}}         
  \frac{d\ka^2}{\ka^2}\frac{\as(\ka^2)}{\pi}                                 
  \ln\left(\frac{e^{\gam_E}\,\nu\,\ka}{2A_{ab}Q}\right).
\end{equation}
Combining together the various pieces and recalling the definition of
$A_{ab}$ in \eqref{eq:A} we find the full PT radiator
\begin{equation}
\label{eq:RadPT-fine}
\cR^{\PT}(\nu,x_1,x_2) \>
=\>C_F\left[\,r_1(\nu,Q_1)+r_2(\nu,Q_2)\,\right]\>
+\>C_A\,r_3(\nu,Q_3)\>,
\end{equation}
where the contribution of the hard parton $\#a$ is given by
\begin{equation}
\label{eq:ra-fine}
  r_a(\nu,Q_a)=2\int_{{Q^2}/{\nu}}^{Q^2}
\frac{d\ka^2}{\ka^2}\frac{\as(\ka^2)}{2\pi}\,
\ln\frac{\xi_a Q_a}{\ka}\>+\>
2 \int_{{Q^2}/{\nu^2}}^{{Q^2}/{\nu}}         
\frac{d\ka^2}{\ka^2}\frac{\as(\ka^2)}{2\pi}                            
\ln\left(\frac{e^{\gam_E}\,\nu\,\ka\, Q_a}{2x_a Q^2}\right)\>.
\end{equation}
Here the hard scales $Q_a$ are
\begin{equation}
  \label{eq:Qa}
\begin{split}
  &Q^2_1=Q^2_2=2(P_1P_2)=Q^2(1-x_3)\>,\qquad \\&
  Q^2_3=\frac{2(P_1P_3)(P_3P_2)}{(P_1P_2)}
  =Q^2\frac{(1-x_1)(1-x_2)}{(1-x_3)}\>.
\end{split}
\end{equation}
In \eqref{eq:ra-fine} we have included into the first term the
rescaling factor $\xi_a$
\begin{equation}
  \label{eq:xi}
  \xi_1=\xi_2=e^{-3/4}\>,\qquad  
  \xi_3=e^{-\be_0/4N_c}\>,
\end{equation}
to take into account the hard collinear splittings of the quark and
the gluon.  These constants, as well as the precise expressions for
the geometry dependent scales $Q_a$, are important in order to
incorporate in the final result \eqref{eq:RadPT-fine} for the {\em
radiator}\/ all terms of the order of $\as^n\ln^{n+1}\nu$ and
$\as^n\ln^{n}\nu$.

To first order we find
\begin{equation}
  \label{eq:ra-first}
r_a(\nu,Q_a)=\frac{\as}{2\pi}\>\ln^2 
\left(\frac{e^{\gam_E}\,\nu\,\xi_a\,Q_a^2}{2x_aQ^2}\right) 
+ \ldots
\end{equation}
This result corresponds, at DL level, to a half of the DL radiator in
the $T_m$ case \cite{Kout}.  The relative factor $1/2$ is due to the
fact that in the $T_m$ case the contribution to the observable is
linear in the angle with respect to the event plane, while it is
quadratic in the $D$ case.

The first order contribution to the PT part of the distribution
\eqref{eq:Sigma-fact} is obtained by performing the Mellin integral in
\eqref{eq:S}. We find
\begin{equation}
  \label{eq:Sigma-first}
\frac{d\Sigma^{\PT}(D,y_c)}{d\ln D}= \frac{\as}{2\pi} \left\{
2C_T\,\ln\frac{1}{D}+\bar G_{11}(y_c)\right\}+\> \cO{\as^2}, 
\end{equation}
with $C_T=2C_F+C_A$ the sum of the colour factors --- the total colour
charge of the hard $\qq,g$ system.  Here $\bar G_{11}$ is given by
\begin{equation}
  \label{eq:barG11}
\bar G_{11}(y_c)=\int dx_1dx_2\Theta(y_3-y_c)
\frac{d\sigma^{(0)}}{\sigma(y_c)dx_1dx_2}\,
G_{11}(x_1,x_2)\>,
\end{equation}
with 
\begin{equation}
  \label{eq:G11}
G_{11}(x_1,x_2)=2C_T\,\ln 27\lam_1\lam_2\>+\> 
2C_F\left[\ln \frac{\xi_1\,Q_1^2}{2x_1Q^2}
      +  \ln \frac{\xi_2\,Q_2^2}{2x_2Q^2}\right]+
2C_A\,   \ln \frac{\xi_3\,Q_3^2}{2x_3Q^2}\>.
\end{equation}
We have checked that this function correctly describes the first
non-logarithmic correction by comparing with the result of the
numerical program EVENT2 \cite{EVENT2} for $y_c=0.025,\, 0.05$ and
$0.1$.

\section{NP calculation \label{sec:NP}}
We follow the usual procedure \cite{Milan} for computing the leading
NP corrections, including two-loop order to take into account the
non-inclusiveness of jet observables.  We start from the PT expression
\eqref{eq:rabcm} and then:
\begin{itemize}
\item we represent the coupling by the dispersive integral \cite{DMW}
\begin{equation} \label{eq:dispersive}
\frac{\as(\ka^2)}{\ka^2}=\int_0^{\infty}
\frac{dm^2\,\aef(m^2)}{(\ka^2+m^2)^2}\>; \end{equation}
\item 
we substitute $\sqrt{\ka^2+m^2}$ for 
the momentum $\ka$ in the source in \eqref{eq:rabcm};
\item 
we take the leading part of the integrand for small $\ka$ and $m$ by
linearising the source \begin{equation}
\label{eq:liear}
\left[1-e^{-\nu\frac{\sqrt{\ka^2+m^2}}{Q}\tau}\right]\>\to\>
\nu\frac{\sqrt{\ka^2+m^2}}{Q}\tau\>,
\end{equation} 
since the NP part of the ``effective coupling'' $\de\aef(m)$ has a
support only at small $m$;
\item 
   we multiply this expression by the Milan factor $\cM$, computed at
   two-loop order, to take into account effects of non-inclusiveness
   of jet observables.
\end{itemize}

For the $\{ab\}$-dipole contribution this procedure gives
\begin{equation}
  \label{eq:drNP'}
\begin{split}  
\de r_{ab}(\nu,x_1,x_2)
= \nu\cM\! \int \! dm^2\frac{\de\aef(m)}{\pi}\frac{-d}{dm^2}\!
\int\frac{d^2\ka}{\pi(\ka^2+m^2)}
\frac{\sqrt{\ka^2+m^2}}{Q}
\int_{-\infty}^{\infty}  d\eta \,\tau(\eta)\>,
\end{split}
\end{equation}
where $\tau$ is the function of $x_1,x_2$ defined in \eqref{eq:tau}.
It is related with the ratio of transverse momentum to energy and
therefore decreases exponentially in rapidity.  This allows us to
extend to infinity the $\eta$ integrals by setting $\eta_M=\infty$.

Here lies the main difference with the $T_m$ case in which the
observable was uniform in rapidity so that the corresponding NP
radiator involved a divergent rapidity integral. There one had to keep
the hard parton recoil momentum which provided an effective cutoff to
the rapidity integral and resulted in a $\log$--enhanced NP
contribution.

We find the leading NP correction to the $\{ab\}$-radiator
\begin{equation}
  \label{eq:drNP}
\de r_{ab}(\nu,x_1,x_2)= \nu\,\frac{a^{\NP}}{Q} g_{ab}(x_1,x_2)\>,
\end{equation}
where $g_{ab}$ is the geometry dependent function 
\begin{equation}
  \label{eq:gab}
g_{ab}(x_1,x_2)\equiv 
\int_{-\pi}^{\pi}\frac{d\phi}{2\pi}\int_{-\infty}^{\infty}
\frac{d\eta\>\sin^2\phi}{A_{ab}\cosh\eta-\cos\phi\sqrt{A_{ab}^2-1}}\>,
\end{equation}
and  $a^{\NP}$ is the NP parameter
\begin{equation}
  \label{eq:a}
  a^{\NP}=2\cM\int dm\frac{\de\aef(m)}{\pi}\>.
\end{equation}
Combining all pieces we obtain the $1/Q$ correction to the radiator
\begin{equation}
\label{eq:RadNP-fine}
\de\cR(\nu)=\nu\, \frac{a^{\NP}}{Q}\Delta(x_1,x_2)\>,
\qquad
\Delta=\frac{N_c}{2}\left(g_{13}+g_{23}-\frac{1}{N_c^2}\,g_{12}\right)\>.
\end{equation}
After merging the PT and NP corrections we can write the NP parameter
in the form
\begin{equation}
  \label{eq:al0}
  a^{\NP}=\frac{4\mu_I} {\pi^2}\cM
\left\{\al_0(\mu_I)-\bar{\al}_s-\be_0\,\frac{\bar{\al}_s^2}{2\pi}
\left(\ln\frac{Q}{\mu_I}+\frac{K}{\be_0}+1 \right)\right\}\>,
\qquad \bar{\al}_s=\al_{\MSbar}(Q)\>,
\end{equation}
where
\[
 \al_0(\mu_I) \equiv \frac{1}{\mu_I}\int _0^{\mu_I} \!dk\,\,\as(k^2)\,.
\]
The term proportional to $K$ accounts for the mismatch between the
$\MSbar$ and the physical scheme \cite{CMW} and reads
\begin{equation}
\label{eq:K}  
K=N_c\left(\frac{67}{18}-\frac{\pi^2}{6}\right)-\frac{5n_f}{9}\>.
\end{equation}
For the analytical expression of the Milan factor $\cM$ see
\cite{graham}.  To quantify the parameter $a^{\NP}$ we recall the
expression for the NP shift $\Delta_T$ in the thrust distribution
\cite{Milan},
\begin{equation}
  \label{eq:thrust} \Delta_T=2C_F\frac{a^{\NP}}{Q}\>.
\end{equation}

\section{Final result \label{sec:Final}}
We are now in a position to obtain the full distribution
$\Sigma(D,y_c)$ in \eqref{eq:Sigma-fact} to the standard accuracy.
First we obtain the PT expression for the soft factor
$S^{\PT}(D,x_1,x_2)$ by performing the Mellin integral \eqref{eq:S}
with the PT radiator \eqref{eq:RadPT-fine}. Then, using the exact
result of the $\cO{\as^2}$ matrix element calculation, we compute the
first correction of the coefficient function $C(\as)$ in
\eqref{eq:Sigma-fact}. This allows us to perform the matching of the
resummed and the exact result to this order. Finally, we include the
leading NP correction from \eqref{eq:RadNP-fine}.

\subsection{PT resummed distribution}
The PT contribution is given by
\begin{equation}
  \label{eq:SPT}
\begin{split}
&S^{\PT}(D,x_1,x_2)=\int \frac{d\nu\,e^{\,\nu \lam_3}}{2\pi i \nu}\>
e^{-\cR^{\PT}(\nu,x_1,x_2)} 
\simeq \frac{e^{-\cR^{\PT}\left(\lam_3^{-1}\!,x_1,x_2\right)}}
{\Gamma(1+\cR')}\>,
\\& 
\cR'\equiv -D\partial_{D}\, 
\cR^{\PT}\left(D^{-1}\!,x_1,x_2\right), 
\end{split}
\end{equation}
This evaluation of the Mellin integral is accurate to SL level, (see
e.g.  \cite{Kout}).  
In the calculation of the logarithmic derivative of $\cR$ the precise
expression for the hard scale $Q_a$ in $r_{a}(\nu,Q_a)$ is beyond SL
accuracy as long as it is of order $Q$.  We can use then $Q$ as a
common hard scale in the SL functions $r'$, and substitute the
logarithmic derivative of the radiator $\cR'$ with
\begin{equation}
\label{eq:R'T}
\cR'_{T}\equiv C_T\,\int_{(DQ)^2}^{DQ^2}\frac{d\ka^2}{\ka^2}
\frac{\as(\ka^2)}{\pi}\>,
\end{equation}
where we have explicitly used the expression \eqref{eq:rab-fine}.

Again, to SL accuracy, we can expand the radiator in the exponent of
\eqref{eq:SPT} and write
\begin{equation}
  \label{eq:RadPTD} \cR^{\PT}\left(\lam_3^{-1}\!,x_1,x_2\right)=
  \cR^{\PT}\left(D^{-1}\!,x_1,x_2\right)+
\cR'_{T}\cdot \ln (27\lam_1\lam_2)\>,
\end{equation}
where $\cO{\as}$ corrections have been neglected.  We conclude by
giving the SL--accurate expression 
\begin{equation}
  \label{eq:SPT-fine}
\begin{split}
S^{\PT}(D,x_1,x_2)=
e^{-\cR^{\PT}\left(D^{-1}, x_1, x_2\right)}\cdot
\frac{e^{-\cR'_{T}\,\ln (27\lam_1\lam_2)}}
{\Gamma(1+\cR'_{T})}\>.
\end{split}
\end{equation}
Using the expression of the PT radiator given in Appendix~\ref{App:PT}
and $\cR'_T$ given in \eqref{eq:R'T}, this distribution can be written
as
\begin{equation}
  \label{eq:SigM}
\ln S^{\PT}(D,x_1,x_2)=L\,g_1(\bar\al\,L)+g_2(\bar\al\,L,x_1,x_2)\>,\quad
\bar\al=\al_{\MSbar}(Q)\>,\quad L=-\ln D\>,
\end{equation}
where only the logarithmic terms in $D$ were kept, while the finite
corrections $\cO{\as}$ dropped. This precaution is necessary in order
to properly set up the procedure for matching the resummed expression
with the exact fixed order result. The DL contribution, the first term
in \eqref{eq:SigM}, does not depend on $x_1,x_2$.  Therefore the
dependence on the event geometry emerges at the level of subleading SL
effects.  In the first order we recover from \eqref{eq:SigM} the
result already obtained in \eqref{eq:Sigma-first}.

Finally, the resummed PT distribution is given by
\begin{equation}
  \label{eq:SigmaPT}
  \Sigma^{\PT}(D,y_c)=
\int dx_1dx_2\Theta(y_3-y_c)\frac{d\sigma^{(0)}}{\sigma(y_c)dx_1dx_2}
\>S^{\PT}(D,x_1,x_2)\>.
\end{equation}

\subsection{Matching resummed and fixed-order prediction}
Here we consider the matching with the exact result of order $\as^2$,
which allows us to compute the first order contribution to the
coefficient function $C(\as)$ in \eqref{eq:Sigma-fact}.  The exact
first order result reads
\begin{equation}
  \label{eq:exact1}
\Sigma_{\exact}^{\PT}(D,y_c)=1+\frac{\as}{2\pi}
\left(-C_T \ln^2\frac{1}{D} 
-\bar G_{11}(y_c)\,\ln\frac{1}{D} +c_1(D,y_c)\right)+\cO{\as^2}\>,
\end{equation}
where $\bar G_{11}(y_c)$ is defined in \eqref{eq:barG11} and
$c_1(D,y_c)$ is a function (regular at $D=0$) which we calculate
numerically by using the four--parton numerical program EVENT2
\cite{EVENT2}.  This gives
\begin{equation}
  \label{eq:C}
  C(\as)=1+\frac{\as}{2\pi}c_1(D,y_c)+\ldots
\end{equation}
In fig.~\ref{fig:c1} we report $c_1(D,y_c)$ as a function of $D$ for
three values of $y_c$.

\EPSFIGURE[ht]{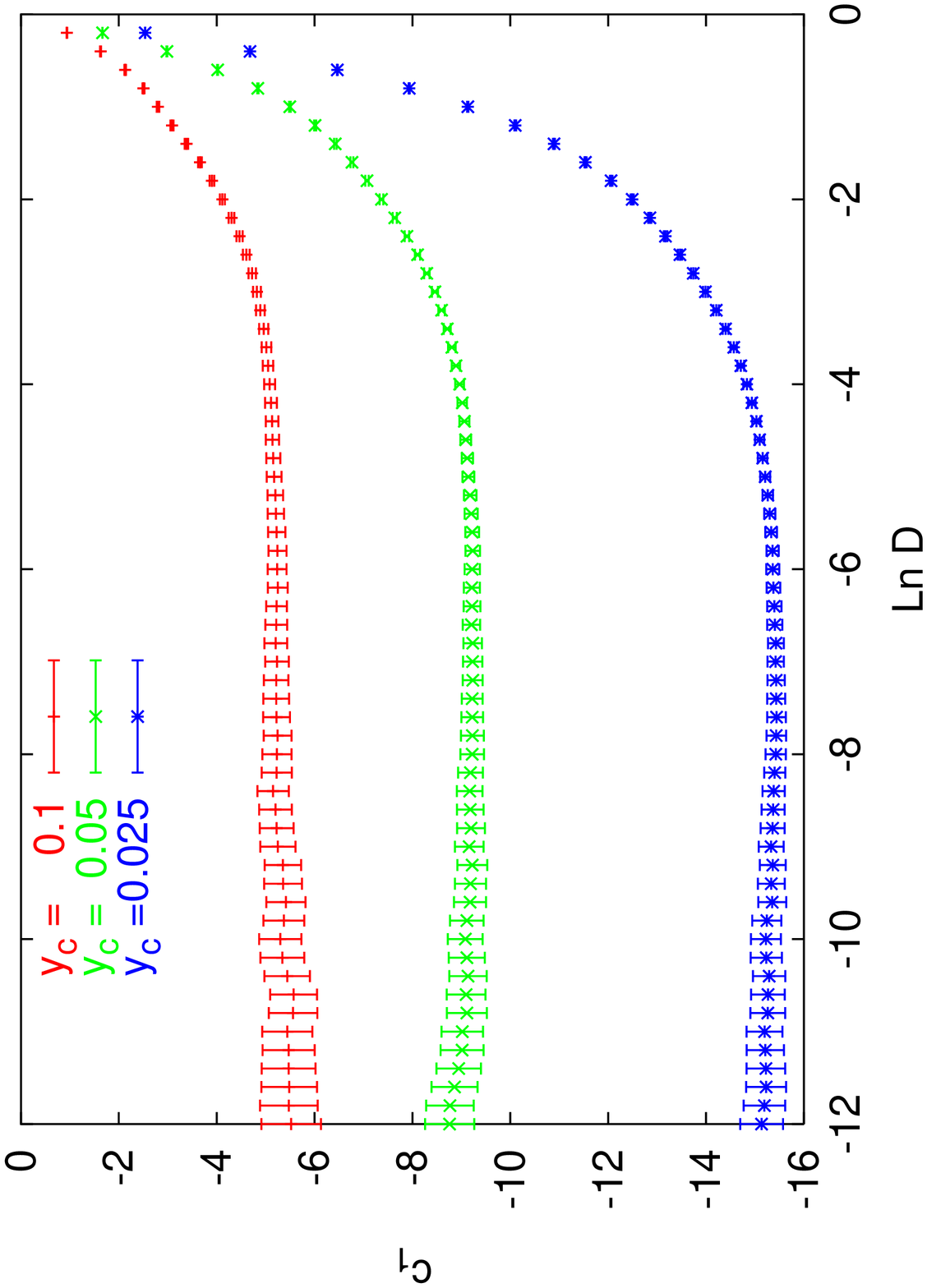,width=0.5\textwidth,angle=270}
{The first order coefficient function for three
        different values of $y_c$.
\label{fig:c1}}

We define the three matched expressions:
\begin{itemize}
\item Log-R matching: 
  \begin{equation}
    \label{eq:mat-logR}
\Sigma^{\PT}_{\mat}(D,y_c)\equiv 
e^{\frac{\as}{2\pi}\,c_1(D,y_c)}\cdot \Sigma^{\PT}(D,y_c)\>, 
  \end{equation}
\item R-matching: 
  \begin{equation}
    \label{eq:mat-R}
\begin{split}
\Sigma^{\PT}_{\mat}(D,y_c)\equiv\left(1+\frac{\as}{2\pi}c_1(0,y_c)\right)\cdot
\Sigma^{\PT}(D,y_c)\> 
+\frac{\as}{2\pi}\,\left(c_1(D,y_c)-c_1(0,y_c)\right)\>,
\end{split}
\end{equation}
\item modified R-matching: 
  \begin{equation}
    \label{eq:mat-MR}
    \Sigma^{\PT}_{\mat}(D,y_c)\equiv
\left(1+\frac{\as}{2\pi}\,c_1(D,y_c)\right)\cdot
\Sigma^{\PT}(D,y_c)\>.
  \end{equation}
\end{itemize}
It is easy to see that each of these expressions reproduces the first
order result \eqref{eq:exact1} and accounts for all terms of order
$\as^n\,\ln^mD$ with $m\ge 2n\!-\!2$.
To obtain also the $\as^n\,\ln^{2n-3}D$ terms one needs a second order
matching ($\cO{\as^3}$ exact matrix element calculation) which
requires five-parton generators \cite{DEBRECEN}, \cite{5partons}.  The
matching procedure becomes more involved and will be discussed in
\cite{AGG}.

\subsection{Including NP corrections}

We now consider the distribution with the NP corrections included.
Using the expression \eqref{eq:RadNP-fine} for the NP radiator, we
have
\begin{equation}
  \label{eq:S-fine}
\begin{split}
S(D,x_1,x_2)= \int \frac{d\nu\,e^{\,\nu \lam_3}}{2\pi i\nu}\>
e^{-\left\{\nu\,\frac{a^{\NP}}{Q}\,\Delta
          +\cR^{\PT}(\nu)\right\}}
=
S^{\PT}\left(D\!-\!\de D,\,x_1,x_2\right)\>,
\end{split}
\end{equation}
where 
\begin{equation}
  \label{eq:DeltaD}
\de D=\frac{a^{\NP}}{Q}\cdot\cD(x_1,x_2)\>,\quad 
\cD(x_1,x_2)=27\lam_1\lam_2\,\Delta(x_1,x_2)\>.
\end{equation}
As for other shape observables, we have embodied the leading NP
correction as a $1/Q$ shift of the argument of the PT distribution.
The magnitude of the shift is determined by the product of the
universal NP parameter and the geometry dependent function
$\cD(x_1,x_2)$ with $\Delta(x_1,x_2)$ given in \eqref{eq:RadNP-fine}.

The final expression that includes the NP correction is given (for
instance, in the Log-R matching scheme) by
\begin{equation}
  \label{eq:Sigma-fine}
  \Sigma(D,y_c)=\int dx_1dx_2\Theta(y_3-y_c)
\frac{d\sigma^{(0)}}{\sigma(y_c)dx_1dx_2}\,
e^{\frac{\as}{2\pi}\,c_1(D',y_c)}\cdot 
S^{\PT}(D',x_1,x_2)\>,
\end{equation}
where $D'$ is the shifted variable
\begin{equation}
  \label{eq:D'}
  \frac{1}{D'}=\frac{1}{D-\de D}-\frac{1}{1-\de D}+1\>.
\end{equation}
The two last terms in \eqref{eq:D'} are relevant only at large $D$ and
have been added as to ensure the correct normalisation of the
distribution, $D'=1$ at $D=1$.

\subsection{Mean value}
We can also consider the mean value of $D$ at fixed $y_c$ defined as
\begin{equation}
  \label{eq:meandef}
  \VEV{D}=\int_0^1 dD D\>\frac{d\Sigma(D,y_c)}{dD}\>.
\end{equation}
The integral is dominated by the region of finite $D$ where the
secondary partons are all hard, and out of the event plane, so that no
logarithmic enhancements are involved here.  The PT contribution $
\VEV{D}^{\PT}$ should be then obtained from a fixed order calculation.
We have to add to it, however, a NP correction
\begin{equation}
  \label{eq:D}
  \VEV{D}= \VEV{D}^{\PT} + \VEV{D}^{\NP}\>.
\end{equation}
The NP correction is determined by soft (small transverse momentum)
partons and can be obtained from the present analysis, cf.\ 
\cite{Kout}. It is simply related with the average value of the shift
as follows
\begin{equation}
  \label{eq:meanNP}
  \VEV{D}^{\NP}=\frac{a^{\NP}}{Q}\,
\int dx_1dx_2\Theta(y_3-y_c)
  \frac{d\sigma^{(0)}}{\sigma(y_c)dx_1dx_2}\cdot \cD(x_1,x_2)\>,
\end{equation}
with $\cD(x_1,x_2)$ given in \eqref{eq:DeltaD}\footnote{The expression
  \eqref{eq:meanNP} has been checked by G.P. Salam and Z. Tr\'ocsanyi
  using their (unpublished) numerical program.}.

\section{Numerical analysis \label{sec:Numerical}}

We report here the numerical results for some typical values of $y_c$.
Data are not yet available.  The results depend on the two parameters
$\al_{\MSbar}(M_Z)$ and $\al_0(\mu_I)$ (with $\mu_I=2\GeV$) which
values we fix in the range determined by the $2$-jet shape
analysis, see for instance ~\cite{GZ}.

In fig.~\ref{fig:mean-Q} we plot, as a function of $Q$, the mean value
$\VEV{D}$ given in \eqref{eq:meandef} for $y_c=0.05$. The leading
order PT contribution is obtained from EVENT2~\cite{EVENT2}. The
next-to-leading order PT contribution is obtained from
DEBRECEN~\cite{DEBRECEN}. The NP contribution is given by
\eqref{eq:meanNP}.

\EPSFIGURE[ht]{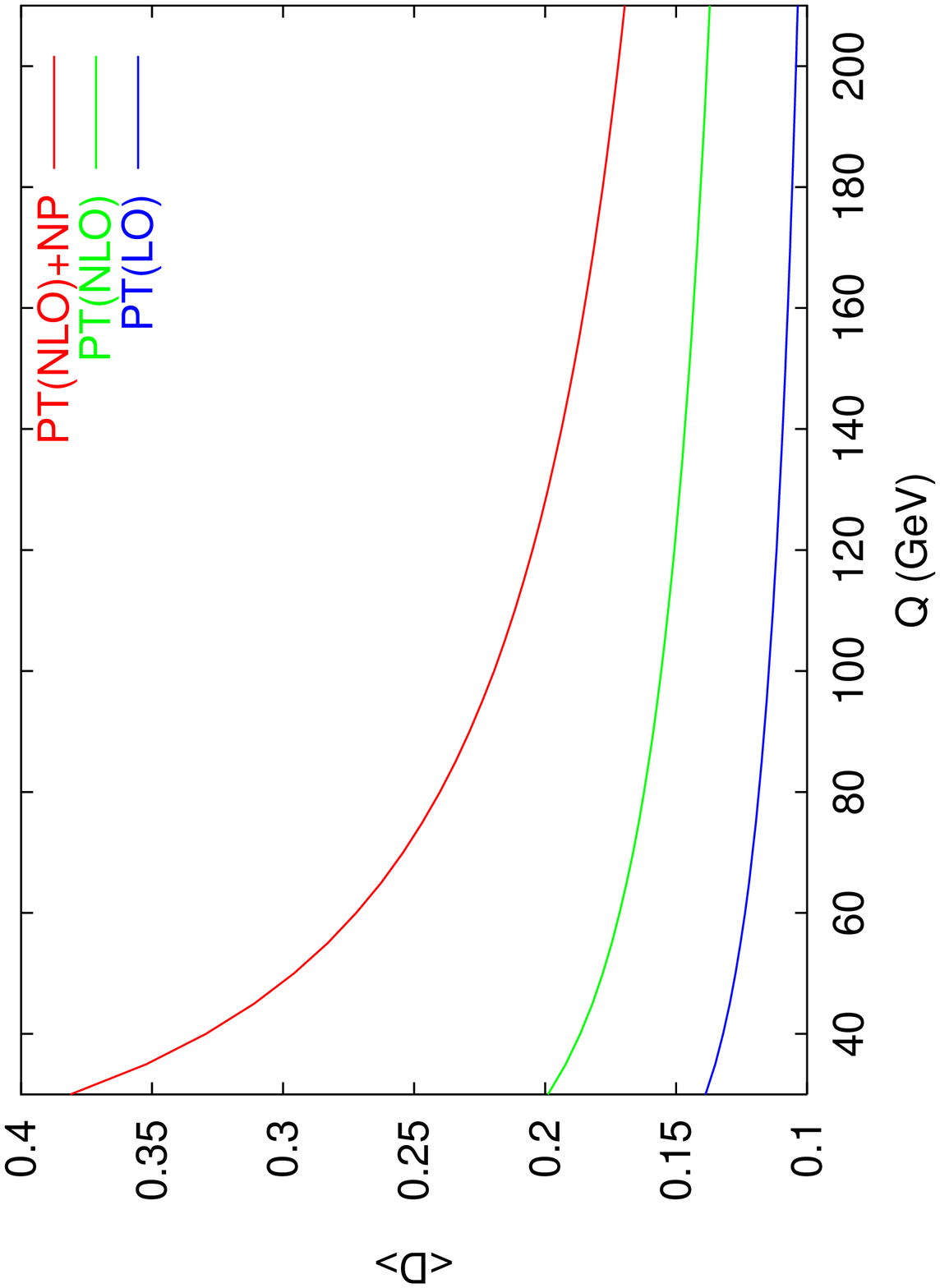,width=0.5\textwidth,angle=270}
{The mean value $\VEV{D}$ as a function of $Q$ for
        $y_c=0.05$. Here we have taken $\al_{\MSbar}(M_Z)=0.118$ 
        and the NP parameter $\al_0(2\GeV)=0.44$.
\label{fig:mean-Q}}
We see that the NP correction is large up to LEP-II energies.
Actually, the coefficient $\Delta(x_1,x_2)$ of the NP correction is
large due to the fact that one of the three radiating hard partons is
the gluon whose contribution to the NP radiation is proportional to
its large colour charge, see \eqref{eq:RadNP-fine}.

In fig.~\ref{fig:mean-yc} we plot the mean value $\VEV{D}$ given in
\eqref{eq:meandef} for three different values of $y_c$ at $Q=91.2\GeV$.
The mean value decreases with $y_c$ decreasing.  This is expected
since decreasing $y_c$ one includes configurations in which the hard
gluon become close to one of the quarks, so that the independent
radiation off the gluon gets suppressed.
\EPSFIGURE[ht]{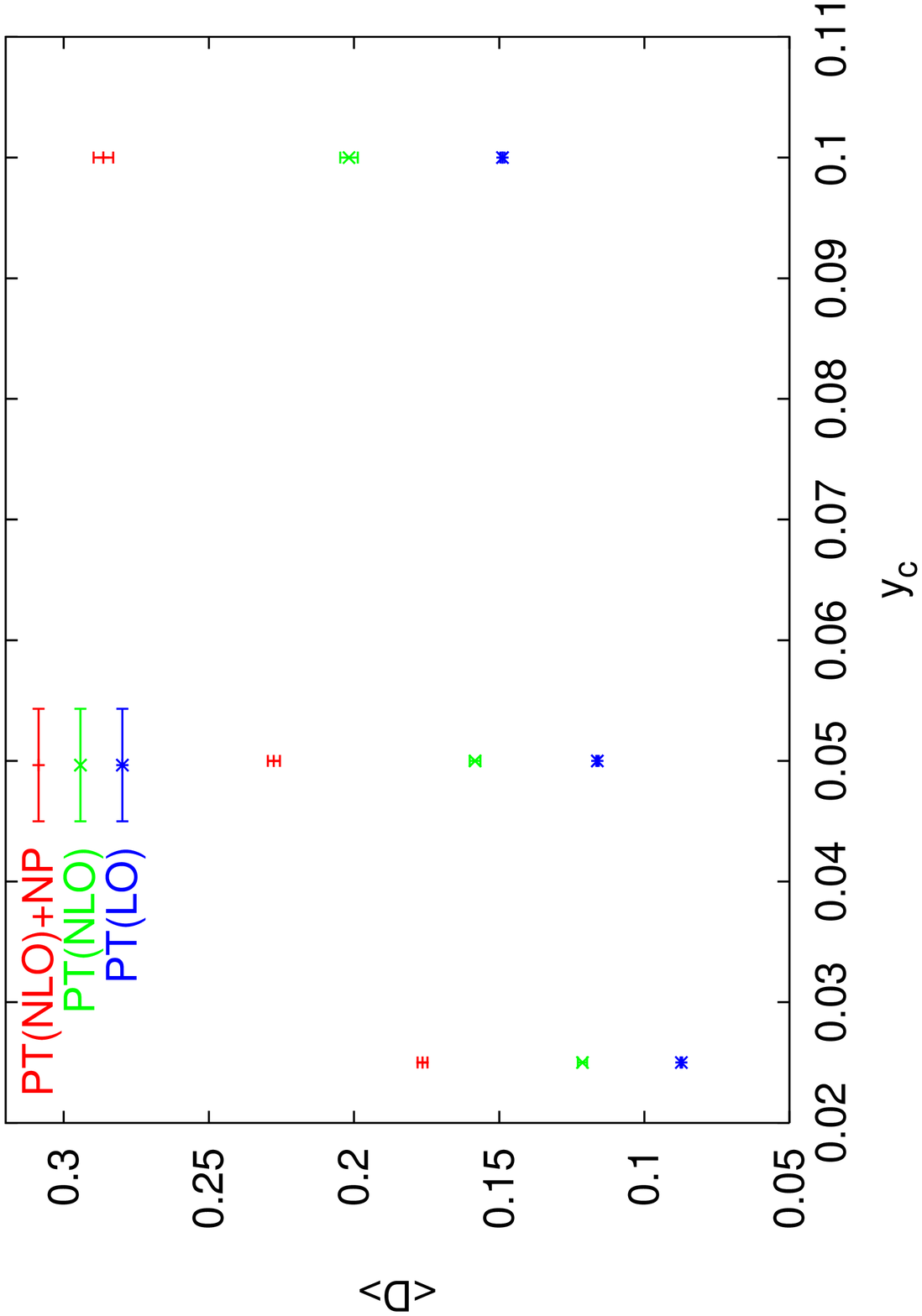,width=0.5\textwidth,angle=270} {The
  mean value $\VEV{D}$ as a function of $y_c$ for $Q=91.2\GeV$ for
  $\al_{\MSbar}(M_Z)$ and $\al_0(\mu_I)$ taken at the values of the
  previous figure.
\label{fig:mean-yc}}
In figs.~\ref{fig:Dst0.025}, \ref{fig:Dst0.05} and \ref{fig:Dst0.1} we
plot the distributions $\Sigma(D,y_c)$ for three values of $y_c$ for
$Q=91.2\GeV$.  The PT distribution is given by the first order Log-R
matched expression in \eqref{eq:mat-logR}. (The second order matching
will be presented elsewhere \cite{AGG}.) The full distribution is
given by the expression \eqref{eq:Sigma-fine}.  The PT peak is shifted
to the right by an amount of order of the NP part of the mean values,
see fig.~\ref{fig:mean-yc}. The shift of the PT distribution is not
uniform since the $x_1,x_2$--dependent shift has been averaged over
the Born distribution.

\EPSFIGURE[ht]{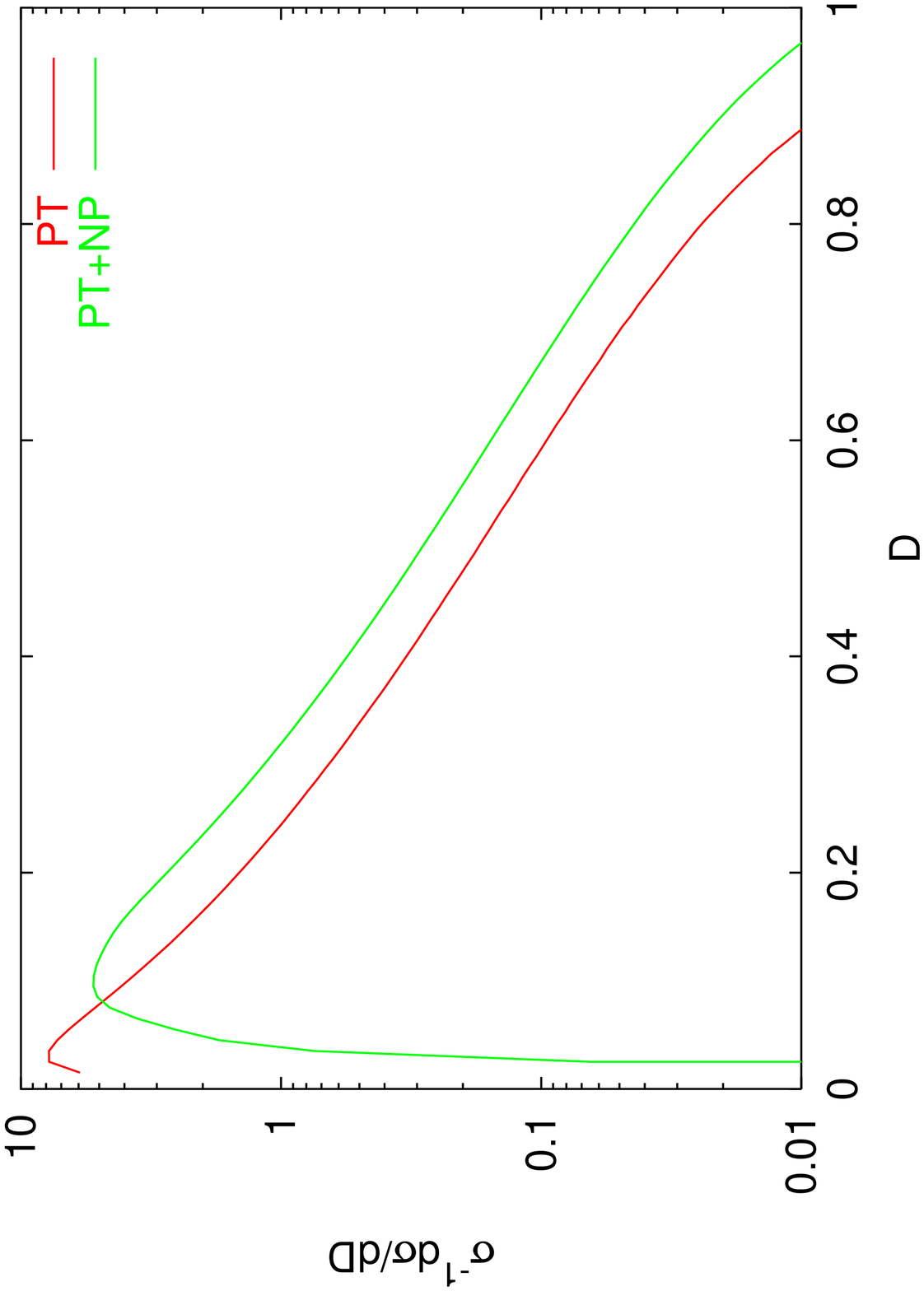,width=0.5\textwidth,angle=270}
{The distribution for $Q=91.2\GeV$ and $y_c=0.025$ for 
        $\al_{\MSbar}(M_Z)=0.118$ and the NP parameter
        $\al_0(\mu_I)=0.52$ at $\mu_I=2\GeV$.
\label{fig:Dst0.025}}
\EPSFIGURE[ht]{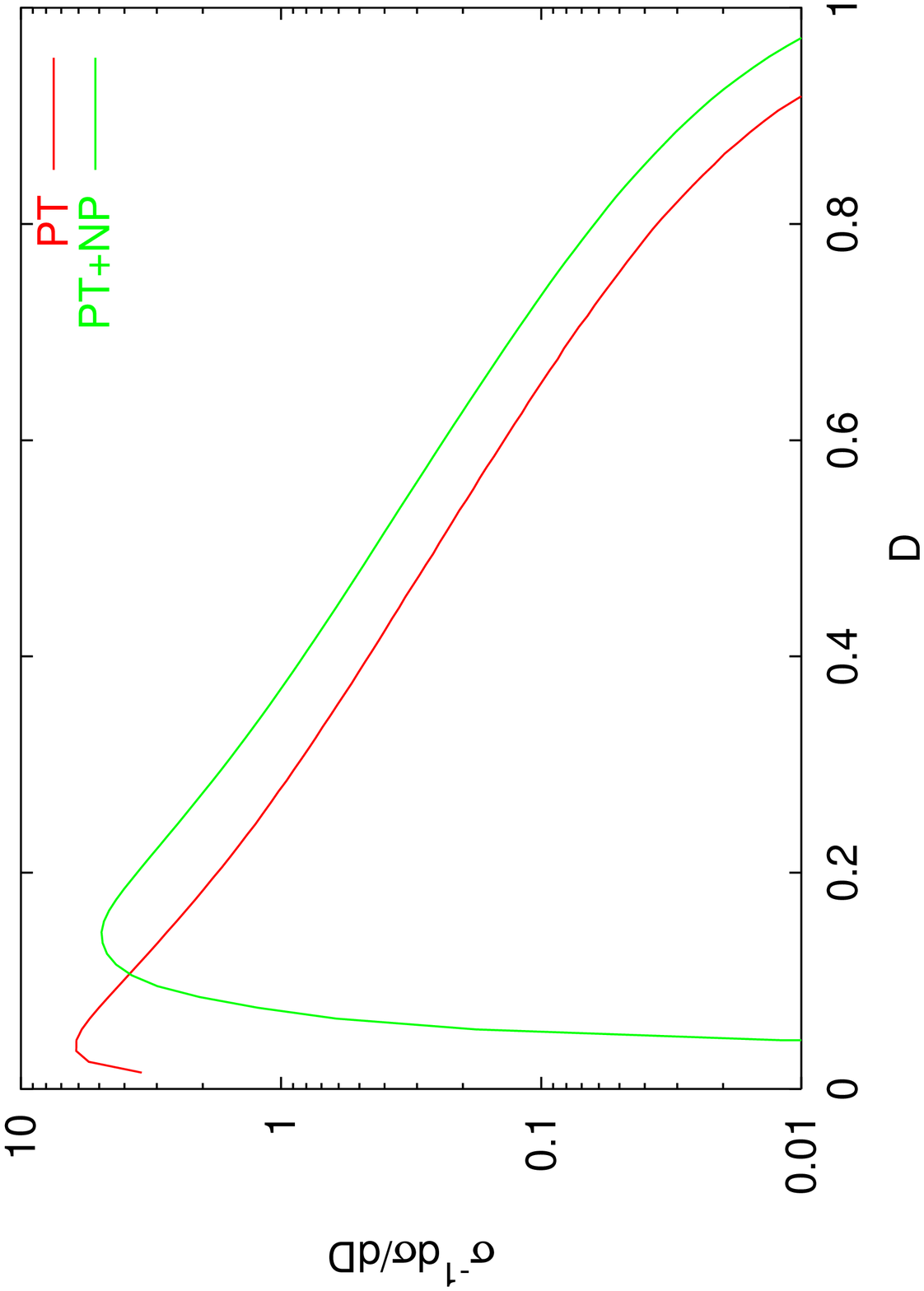,width=0.5\textwidth,angle=270}
{The distribution for $Q=91.2\GeV$ and $y_c=0.5$ for 
        $\al_{\MSbar}(M_Z)=0.118$ and the NP parameter
        $\al_0(\mu_I)=0.52$ at $\mu_I=2\GeV$.
\label{fig:Dst0.05}}
\EPSFIGURE[ht]{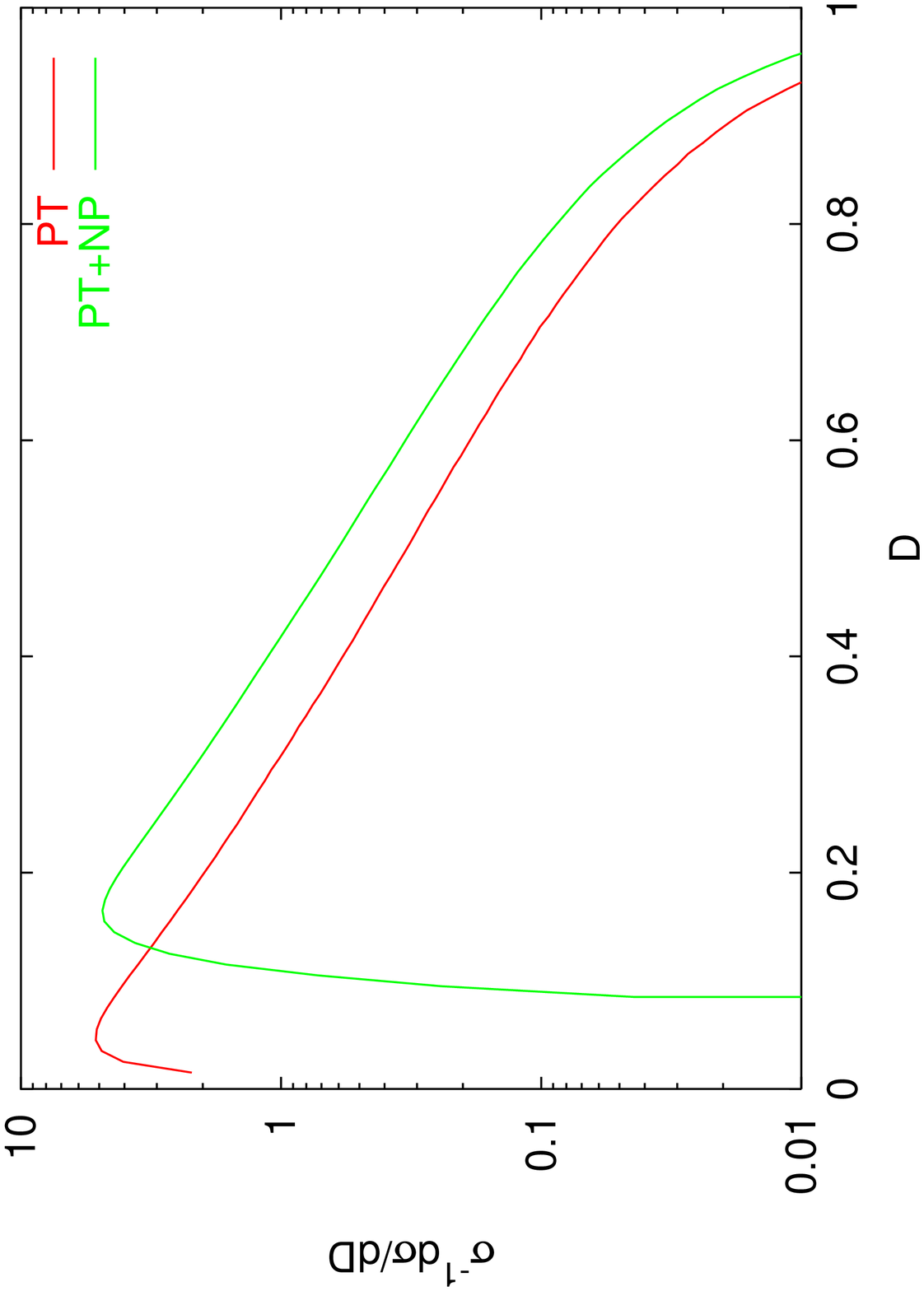,width=0.5\textwidth,angle=270}
{The distribution for $Q=91.2\GeV$ and $y_c=0.1$ for 
        $\al_{\MSbar}(M_Z)=0.118$ and the NP parameter
        $\al_0(\mu_I)=0.52$ at $\mu_I=2\GeV$.
\label{fig:Dst0.1}}

\section{Discussion and conclusion \label{sec:Discussion}}
The aim of the present study is the understanding of the structure of
multi-jet events in $\ee$.  To master this field is important,
especially, for hadron-hadron collisions where the QCD events involve
more than two jets.

This is the second near-to-planar three-jet shape variable, after
$T_m$, studied at an accuracy sufficient to make quantitative
predictions (SL resummation, matching with fixed order, NP corrections).
The main difference between the $D$-parameter and $T_m$ is the
dependence of the observable on particle rapidity. Actually, $T_m$ is
uniform in rapidity, while $D$ is exponentially dumped. This feature
has important consequences both at PT and NP level.

On the PT side, the hard parton recoil explicitly enters the
observable in the $T_m$ case. On the contrary, in the $D$ case it
gives rise to corrections beyond SL accuracy. This in turn implies
that the PT $D$ distribution depends only on the geometry of the event
(the angles between jets) and not on its colour configuration.

In what concerns NP corrections, one finds that in the $D$ case only
soft radiation at large angles contributes to the NP shift. As a
consequence, the shift is not logarithmically enhanced, as was the
case of $T_m$.  This is the same difference that occurs between
$1-T,C,M^2/Q^2$ and $B$ jet shapes in the two-jet case.

We observe that the shift for the $D$ parameter is larger than is
typical for 2-jet observables. This is explained by the fact that in
3-jet events we have three hard radiating partons whose total colour
charge $2C_F + C_A$ is significantly larger than $2C_F$, the total
colour charge of a two-jet system.  As a consequence, while for 2-jet
observables higher order NP effects come into play below the peak of
the distribution, for the $D$ parameter they are already relevant in
the proximity of the peak.  This calls for a deeper analysis that
would address higher powers in $1/Q$, for example, along the lines of
Korchemsky-Sterman approach which was recently developed for some
2-jet observables in \cite{KS}.  The comparison with experimental data
(not yet available) would shed light on this important point.

Near-to-planar 3-jet events provide a new method to measure the
fundamental QCD parameter $\as(M_Z)$ and to have a further test of the
universality of genuine NP effects in jet physics.

\section*{Acknowledgements}
We are grateful to Gavin Salam and Bryan Webber for helpful
discussions and suggestions.  We thank also G\"unther Dissertori and
Oliver Passon for discussions on future analyses of LEP data.

\appendix

\section{Born kinematics \label{App:Born}}
We select the event plane in such a way that the two Born momenta
with the maximum and minimum $x$-component are given by
\begin{equation}
  \label{eq:Pa}
\begin{split}   
  & P_{\max}=E(x_{\max},0,0,x_{\max})\>,\qquad
  P_{\min}=E(x_{\min},0,-T_M,-t_{\min})\>.
\end{split}
\end{equation}
with $Q=2E$ and
\begin{equation}
  \label{eq:xs}
\begin{split}
x_{\max}\!=\!T\>,\quad 
x_{\min}\!=\!1\!-\!\frac{T}{2}(1\!+\!\rho)\>,\quad
t_{\min}\!=\!\frac{T}{2}\!-\!\rho\left(1-\frac{T}{2}\right)\>,\quad
\rho\equiv\sqrt{1\!-\!\frac{T_M^2}{1\!-\!T}}\>,
\end{split}
\end{equation}
so that the thrust major is 
\begin{equation}
\label{eq:TTM}
T_M=\frac{2}{x_{\max}}\sqrt{(1-x_1)(1-x_2)(1-x_3)}\>.
\end{equation}
To define these massless momenta one needs that $T$ and $T_M$ are
restricted to the region
\begin{equation}
  \label{eq:kinB}
\frac{2(1-T)\sqrt{2T-1}}{T}<T_M<\sqrt{1-T}\>.
\end{equation}
In this paper we consider $T,T_M$ restricted to this region.

\section{The PT radiator\label{App:PT}}
Here we compute the radiator component $r_{ab}(\nu)$ given in
\eqref{eq:rabcm}. We change integration variable from $\eta$ to $\tau$
with the Jacobian
\begin{equation}
  \label{eq:Jacobian}
  d\eta=-\frac{d\tau}{\tau}\,K(\tau)\>,\qquad  
K(\tau)=\frac{\sin^2\phi}
{\sqrt{\left(\sin^2\phi+\tau\sqrt{A^2\!-\!1}\,\cos\phi\right)^2
-\tau^2A^2}}\>.
\end{equation}
In this appendix we set $A_{ab}\to A$ and $\eta^0_{ab}\to \eta_0$
defined in \eqref{eq:A}. The $\eta$-integral can be divided into the
positive and negative rapidity regions and one has
\begin{equation}
  \label{eq:rab'}
  r_{ab}(\nu)=r_{ab}^+(\nu)+r_{ab}^-(\nu)\>,
\end{equation}
where
\begin{equation}
\label{eq:rpm}
r_{ab}^\pm(\nu)=2\int_0^{Q}
\frac{d\ka}{\ka}\frac{\as(\ka^2)}{\pi}\int_{-\pi}^{\pi}\frac{d\phi}{2\pi}
\int_{\tau_M^{\pm}}^{\tau_0}\frac{d\tau}{\tau}\,K(\tau)
\left(1-e^{-\nu\frac{\ka}{Q}\tau}\right).
\end{equation}
The limits are for $\eta=\pm\eta_0$ and $\eta=\eta_M$
\begin{equation} 
  \label{eq:taulim}
\tau_0=\frac{\sin^2\phi}{A\!-\!\sqrt{A^2\!-\!1}\,\cos\phi}\>,
\qquad \tau_M^{\pm}=\frac{\sin^2\phi}
{A \cosh(\eta_M\pm\eta_0)\!-\!\sqrt{A^2\!-\!1}\,\cos\phi}\>.
\end{equation}
The limit $\tau_M^{\pm}$ is a function of $\ka$ given, for
small $\ka$, by
\begin{equation}
  \label{eq:tausmall}
  \tau_M^{\pm}(\ka)\simeq\frac{2\sin^2\phi}{A}\frac{\ka}{Q^{\pm}_{ab}}\>,
\qquad Q_{ab}^{\pm}=Q_{ab}e^{\pm\eta_0}\>.
\end{equation}
In \eqref{eq:rpm} we can substitute, to SL accuracy, the factor
$[1-e^{-\nu\frac{\ka}{Q}\,\tau}]$ by the effective cutoff
\eqref{eq:cutoff} and we obtain
\begin{equation}
\label{eq:rpm1}
r_{ab}^\pm(\nu)\simeq 2\int_0^{Q}
\frac{d\ka}{\ka}\frac{\as(\ka^2)}{\pi}\int_{-\pi}^{\pi}\frac{d\phi}{2\pi}
\int_{\tau_M^{\pm}}^{\tau_0}\frac{d\tau}{\tau}\,K(\tau)
\>\vartheta\left(\tau-\frac{Q}{\bnu \ka}\right). 
\end{equation}
In order to reach SL accuracy, we have to take the running coupling at
two loops and include the non-soft part of the parton splitting
function. Before discussing this case, it is instructive to consider
the first order term in $\as=\as(Q)$.

\subsection{First order result}
We change again variables $q=\ka\tau$ and obtain
\begin{equation}
  \label{eq:rabpm}
r_{ab}^{\pm}
= \frac{2\as}{\pi}\int_{{Q}/{\bnu}}^{Q}
\frac{dq}{q}%\left(1-e^{-\nu\frac{q}{Q}}\right)
\int_{-\pi}^{\pi}\frac{d\phi}{2\pi}
\int_{\tau_{\pm}}^{\tau_0}\frac{d\tau}{\tau}\,K(\tau)\>,
\end{equation}
where $\tau_{\pm}$ is the solution of the equation
\begin{equation}
  \label{eq:tauMsmall}
\tau_{\pm}=\tau_M^{\pm}\left(\frac{q}{\tau_{\pm}}\right)
=\sqrt{\frac{2\sin^2\phi}{A}\frac{q}{Q^{\pm}_{ab}}}
+\cO{\frac{q}{Q}}\>,
\end{equation}
with $\tau_M^{\pm}(k)$ the function of $\ka$ given in
\eqref{eq:taulim}.  Again in \eqref{eq:rabpm} we take as upper limit
the hard scale $Q$. For small $q$ the $\tau$ integral is
\[
\int_{\tau_{\pm}}^{\tau_0}\frac{d\tau}{\tau}K(\tau)=
\half\ln\frac{2\,Q^{\pm}_{ab}}{Aq}+\half\ln\sin^2\phi+
\cO{\frac{q}{Q}}\>.
\]
Summing the $r^+_{ab}$ and $r^-_{ab}$
contributions, we get, at SL accuracy, 
\begin{equation}
  \label{eq:rab1}
\begin{split}
r_{ab}(\nu)\simeq
\frac{2\as}{\pi}\int_{Q/\bnu}^{Q}\frac{dq}{q}
\ln\frac{Q_{ab}}{2Aq}\simeq
\frac{\as}{\pi}\,\ln^2
\left(\frac{\bnu Q^2_{ab}}{2\sqrt{x_ax_b}Q^2}\right) \>.
\end{split}
\end{equation}
Finally, using \eqref{eq:RadPT1} we obtain, for fixed $\as$, the
complete radiator $\cR^{\PT}(\nu)$ in \eqref{eq:RadPT-fine}.  The
parton $\#a$ component in \eqref{eq:ra-first} is obtained from
\eqref{eq:rab1} by taking into account the non-soft part of the
splitting function which gives the rescaling factors $\xi_a$ in
\eqref{eq:xi}.

\subsection{SL result}
Here we evaluate $r_{ab}^{\pm}$ in \eqref{eq:rpm1} with the running
coupling at two loops.  
We consider the two regions of the $\tau$-integration: 
$\tau^{\pm}_M>Q/\bnu \ka$ and $\tau^{\pm}_M<Q/\bnu \ka$. 
We then have two contributions
\begin{equation}
\label{eq:r+-1}
\begin{split}
&r_{ab}^{\pm}=2\int_{-\pi}^{\pi}\frac{d\phi}{2\pi}\left\{
\int_{q_{\pm}}^Q
\frac{d\ka}{\ka}\frac{\as(\ka^2)}{\pi}
\int_{\tau^{\pm}_M}^{\tau_0}
\frac{d\tau}{\tau}K(\tau)+
\int_{\frac{Q}{\bnu\tau_0}}^{q_{\pm}}
\frac{d\ka}{\ka}\frac{\as(\ka^2)}{\pi}
\int_{\frac{Q}{\bnu\ka}}^{\tau_0}\frac{d\tau}{\tau}K(\tau) \right\}\\
&\simeq2\int_{-\pi}^{\pi}\frac{d\phi}{2\pi}\left\{
\int_{q_{\pm}}^Q
\frac{d\ka}{\ka}\frac{1}{\pi}
\left(\as(\ka^2)+\as\left(\frac{q_{\pm}^4}{\ka^2}\right)
\right)\ln\frac{Q_{ab}^{\pm}}{\ka}\right\}\,,\qquad
q_{\pm}^2=\frac{QQ_{ab}^{\pm}\,A_{ab}}{\bnu\,2\sin^2\phi}\>,
\end{split}
\end{equation}
where in the second term we made the variable change $\ka\to
q^2_{\pm}/\ka$. The precise expression of the upper limit in the first
term is beyond SL accuracy as long as of order $Q$. In the second line
we have neglected terms beyond SL accuracy ($\as^n\ln^{n-1}\nu$).

Now we can perform the $\phi$-integration observing that only terms
containing one power of $\ln\sin^2\phi$ contribute to the radiator at
SL level.  Therefore we make the substitution
$$
\int_{-\pi}^{\pi}\frac{d\phi}{2\pi}\ln\sin^2\phi=-\ln 4\> \Rightarrow \>
q_{\pm}^2\to q_{\pm}^2=\frac{QQ_{ab}^{\pm}\,2A_{ab}}{\bnu}\>,
$$
In conclusion the radiator may be written as
\begin{equation}
\label{r+-2}
r_{ab}^{\pm}=
2\int_{q^2_{\pm}}^{Q^2}
\frac{d\ka^2}{\ka^2}\frac{\as(\ka^2)}{2\pi}                                
\ln\frac{Q_{ab}^{\pm}}{\ka}+
2\int_{q^4_{\pm}/Q^2}^{q^2_{\pm}}
\frac{d\ka^2}{\ka^2}\frac{\as(\ka^2)}{2\pi}                                
\ln\frac{\bnu\,\ka}{2A_{ab}Q}\>.
\end{equation}
One finds that the dependence on $q_{\pm}$ in the integration
limits cancels out, so that, adding $r_{ab}^+$ and
$r_{ab}^-$, one is left with \eqref{eq:rab-fine}.
Combining together the various pieces and recalling the definition
of $A_{ab}$ in \eqref{eq:A} we find the full radiator
$\cR^{\PT}(\nu)$ in \eqref{eq:RadPT-fine} with the parton
$\#a$ component $r_a$ given by the sum of two terms
\begin{equation}
\label{eq:rU&rL}
\begin{split}
&r_a =r_a^U+r_a^L\>,\\
&r_a^U = 2\int_{{Q^2}/{\nu}}^{Q^2}
\frac{d\ka^2}{\ka^2}\frac{\as(\ka^2)}{2\pi}                                
\ln\frac{\xi_a Q_a}{\ka}\>, \quad
r_a^L =2\int_{{Q^2}/{\nu^2}}^{{Q^2}/{\nu}}         
\frac{d\ka^2}{\ka^2}\frac{\as(\ka^2)}{2\pi}                                
\ln\left(\frac{\bnu\,\ka\,Q_a}{2x_a Q^2}\right).
\end{split}
\end{equation}
The hard scales are given by \eqref{eq:Qa}.  The SL terms coming from
the parton hard collinear splitting are taken into account, see
\cite{Kout} and \cite{TDIS}, simply by rescaling the scale $Q_a$ by
the constants $\xi_a$ in \eqref{eq:xi} in the piece $r_a^U$.  To first
order in $\as$ this expression coincides with the result in
\eqref{eq:ra-first}.

To evaluate the two components $r_a^U,r_a^L$ it is enough to consider
the following expression for the running coupling:
\begin{equation}
\label{eq:alpha}
\as(\mu^2)=\frac{\as}{1-\lambda} \left(1-
\frac{\be_1}{\be_0}\frac{\as}{2\pi}
\frac{\ln(1-\lambda)}{1-\lambda}\right)\>,
\qquad \lambda=\frac{\as\be_0}{4\pi}\ln\frac{Q^2}{\mu^2}\>,
\end{equation}
with 
\begin{equation}
\be_0=\frac{11N_c-2n_f}{3}\,,\qquad
\be_1=\frac{17N_c^2-5N_c\,n_f-3C_F\,n_f}{3}\,,
\end{equation}
and $\as=\as(Q)$ in the physical scheme \cite{CMW} related to the
$\MSbar$ by
\begin{equation}
  \label{eq:K'}
  \as= \bar\as\left(1+\frac{ K\bar\as}{2\pi}+\ldots\right),
\qquad \bar \as =\al_{\MSbar}(Q)\>,
\end{equation}
with $K$ given in \eqref{eq:K}.

We can separate the pieces depending on the geometry. We write 
\begin{equation}
\label{eq:boh}
\begin{split}
r_a^U = r_{DL}^U+r_{SL}^U\cdot \ln\left(\frac{\xi_a Q_a}{Q}\right)^2, \qquad
r_a^L = r_{DL}^L+r_{SL}^L\cdot 
\ln\left(\frac{e^{\gam_E}Q_a}{2x_aQ}\right)^2.
\end{split}
\end{equation}
Here the geometry dependence is explicitly expressed by the scales 
$Q_a$, while the various $r$-functions are evaluated at the common 
hard scale $Q$.
The various functions are
\begin{equation}
\label{eq:PTcomplex}
\begin{split}
 r_{DL}^U =& \int_{\frac{Q^2}{\nu}}^{Q^2} 
\frac{d\ka^2}{\ka^2}\frac{\as(\ka^2)}{2\pi}\ln\frac{Q^2}{\ka^2}
=\frac{8\pi}{\bar\as\be_0^2}\left(-\rho-L(\rho)\right)
-\frac{4\be_1}{\be_0^3}\left(\half L^2(\rho)+
\frac{L(\rho)+\rho}{1-\rho}\right)\\
&+\frac{4K}{\be^2_0}\left(L(\rho)+\frac{\rho}{1-\rho}\right)\>,\\
r_{DL}^L =&\int_{\frac{Q^2}{\nu^2}}^{\frac{Q^2}{\nu}}         
\frac{d\ka^2}{\ka^2}\frac{\as(\ka^2)}{2\pi}\ln\frac{\nu^2\ka^2}{Q^2} 
=\frac{8\pi}{\bar\as\be_0^2}\left(\rho+(2\rho-1)(L(\rho)-L(2\rho))\right)
\\&
-\frac{4\be_1}{\be_0^3}\left(\half (L^2(\rho)-L^2(2\rho))
+(2\rho-1)\left(
\frac{L(2\rho)+2\rho}{1-2\rho}-\frac{L(\rho)+\rho}{1-\rho}
\right)\right)\\
&+\frac{4K}{\be^2_0}\left(L(\rho)-L(2\rho)-\frac{\rho}{1-\rho}\right)\>,\\
r_{SL}^U =&\int_{\frac{Q^2}{\nu}}^{Q^2}\frac{d\ka^2}{\ka^2}
\frac{\as(\ka^2)}{2\pi}=\frac{2}{\be_0}\ln\frac{1}{1\!-\!\rho}\>, 
\quad
r_{SL}^L =\int_{\frac{Q^2}{\nu^2}}^{\frac{Q^2}{\nu}}         
\frac{d\ka^2}{\ka^2}\frac{\as(\ka^2)}{2\pi}=
\frac{2}{\be_0}\ln\frac{1\!-\!\rho}{1\!-\!2\rho}\>.
\end{split}
\end{equation}
where
\begin{equation}
\rho=\bar\as\>\frac{\be_0}{4\pi}\ln\nu\>,\qquad
L(x)=\ln(1-x)\>.
\end{equation}
Notice that the radiator contain only DL and SL terms 
\begin{equation}
  \label{eq:Rad-exp}
  \cR^{\PT}(\nu)=\frac{1}{\as}F_1(\rho)+F_2(\rho)\>.
\end{equation}

\section{Effective cutoff \label{App:cutoff}}
Performing the variable change $q=\ka\tau$ in \eqref{eq:rpm} the
radiator is proportional to
\begin{equation}
\label{eq:uno}
r(Q) = \int_0^{Q} \frac{dq}{q}\left(1-e^{-\nu\frac{q}{Q}}\right)\> I(q)\>,
\qquad I(q)= \int_{-\pi}^{\pi}\frac{d\phi}{2\pi}
\int_{\tau_{\pm}}^{\tau_0}\frac{d\tau}{\tau}\,K(\tau)\>\as(q^2/\tau^2)\>.
\end{equation}
This shows that $I(q)$ has the form 
\begin{equation}
\label{eq:FaL}  
I(q)=F(\as L_q)+\as G(\as L_q)+\ldots\>,
\quad \as=\as(Q)\>,\quad L_q\!=\!\ln\frac{Q}{q}\>,
\end{equation}
where the dots do not contribute at SL level.  We consider first the
contribution from the leading piece of $I(q)$
\begin{equation}
  r'(Q) = \int_0^{Q} \frac{dq}{q}
\left(1-e^{-\nu\frac{q}{Q}}\right)\> F(\as L_q)\>.
\end{equation}
To SL accuracy here we can make the
substitution
\begin{equation}
  \label{eq:cutoff}
  \left(1-e^{-\nu\frac{q}{Q}}\right)\to
\vartheta\left(q-\frac{Q}{\bnu}\right), \qquad \bnu=e^{\gam_E}\nu\>.
\end{equation}
For completeness we prove this well known result (see
\cite{PTstandard}). We can write
\begin{equation}
  \label{eq:CI}
r'(Q)=\int_{\bar q}^{Q}\frac{dq}{q}\,F(\as L_q)+\Delta(\bar q)\>,
\end{equation}
with 
\begin{equation}
  \label{eq:CD}
\begin{split}
\Delta(\bar q)
&=\int_{0}^{\bar q}\frac{dq}{q}
\left(\frac{q}{Q}\right)^{\eps}\,F(\as L_q)-
\int_{0}^{Q}\frac{dq\,e^{-\nu\frac{q}{Q}}}{q}
\left(\frac{q}{Q}\right)^{\eps}F(\as L_q)\\  
&
=F(-\as\partial_{\eps})\cdot
\left\{\int_{0}^{\bar q}\frac{dq}{q}\left(\frac{q}{Q}\right)^{\eps}-
\int_{0}^{Q}\frac{dq\,e^{-\nu\frac{q}{Q}}}{q}
\left(\frac{q}{Q}\right)^{\eps}\right\}.
\end{split}
\end{equation}
for $\eps\to0$.  We can extend the last integral to infinity and get,
up to $e^{-\nu}$ corrections for large $\nu$,
\begin{equation}
  \label{eq:CD1}
\Delta(\bar q)\simeq
F(-\as\partial_{\eps})\cdot\left\{
\frac{1}{\eps}\left(\frac{\bar q}{Q}\right)^{\eps} 
-\frac{\Gamma(1+\eps)}{\eps}\nu^{-\eps}\,\right\}.
\end{equation}
Our aim now is to select $\bar q$ in such a way that $\Delta$ is
beyond SL accuracy.  
Setting $\bar q=Q/\bnu$ we get
\begin{equation}
  \label{eq:CD2}
\Delta(\bar q)=
F(\as L_{\bar q}-\as\partial_{\eps})\cdot f(\eps)\>, \qquad
L_{\bar q}= - \ln \bnu\>.
\end{equation}
where 
\begin{equation}
  \label{eq:gamma}
f(\eps)=\frac{1-\Gamma(1+\eps)\,e^{\eps\gam_E}}{\eps}=\cO{\eps}\>.  
\end{equation}
We conclude then that
\begin{equation}
  \label{eq:CD3}
  \Delta(\bar q)\sim \as\,F'(\as L_{\bar q})\>,
\qquad F'(x)=\partial_x\,F(x)\>,
\end{equation}
which is negligible within SL accuracy.
We conclude then, to SL accuracy, 
\begin{equation}
  \label{eq:CII}
r'(Q) \simeq \int_{Q/\bnu}^{Q}\frac{dq}{q}\,F(\as L_q)\>.
\end{equation}
For the second term $\as G(\as L_q)$ in \eqref{eq:FaL} the analysis is
simpler. Since it gives only a SL contribution the lower scale can be
taken at any value of the order $Q/\nu$. We then conclude that 
to compute the radiator to SL accuracy we can use the cutoff 
substitution \eqref{eq:cutoff}.

\end{document}